\documentclass[twocolumn,prl,aps]{revtex4}
\usepackage[dvips]{graphicx}
\usepackage{amsmath}
\usepackage{amsfonts}
\usepackage{amssymb}

\newcommand{\ar}{\begin{eqnarray}}
\newcommand{\br}{\end{eqnarray}}
\newcommand{\ica}{i_{Ca}}

\newcommand{\kp}{k_{+}}
\newcommand{\km}{k_{-}}

\newcommand{\fnI}{\footnote{The Goldman-Hodgkin-Katz
equation gives, for voltage $V$ (mV), $i_{Ca} = P_{Ca} \phi
(\beta_{Ca} C_{ext} -
e^{\phi}c)/(e^{\phi}-1)\,\mathrm{fmol\,\,s^{-1}}$ where
$\phi=2VF/RT$ with permeability $P_{Ca}=0.913\,\mathrm{\mu m^3 \,
\mu s^{-1}}$, $\beta_{Ca}=0.341$, extracellular [Ca]
$C_{ext}=2000\,\mathrm{\mu M}$, $F=96.5\,\mathrm{C \, mmol^{-1}}$,
$R=8.314\,\mathrm{J \, M^{-1} \, K^{-1}}$, $T=310\,\mathrm{K}$.
Rates for the LCC Markov scheme (Fig. \ref{fig1}) are
$\alpha_1(V)=(1+e^{(V-2)/7}) \,\mathrm{ms^{-1}}$,
$\beta_1(V)=2.35\,\mathrm{ms^{-1}}$, $\alpha=1/9\,\mathrm{ms^{-1}}$
and $\beta=1\,\mathrm{ms^{-1}}$. Other physiological parameters are
$\tau=4.4\,\mathrm{\mu s}$, $v=1.26\cdot 10^{-3}\,\mathrm{\mu
m^{3}}$, $c_{sr}=1000\,\mathrm{\mu M}$, $c_o=0.1\,\mu M$,
$\kp=0.0005\,\mu M^{-2}\,ms^{-1}$, and $\km=2\,ms^{-1}$} }

\begin{document}

\title{Macroscopic consequences of calcium signaling in microdomains:\\
A first passage time approach}

\author{Robert Rovetti$^{1}$, Kunal K. Das$^{2}$, Alan Garfinkel$^{3}$,
and Yohannes Shiferaw$^{4}$ } \affiliation {$^{1}$Department of
Biomathematics, University of California, Los Angeles, California
90095} \affiliation {$^{2}$ Department of Physics, Fordham
University, Bronx NY 10458} \affiliation {$^{3}$ Department of
Medicine, University of California, Los Angeles, California 90095} \affiliation {$^{4}$ Department of Physics and Astronomy,
California State University, Northridge, California 91330}

\date{\today}

\begin{abstract}

Calcium (Ca) plays an important role in regulating various cellular
processes. In a variety of cell types, Ca signaling occurs within
microdomains where channels deliver localized pulses of Ca which
activate a nearby collection of Ca-sensitive receptors.  The small
number of channels involved ensures that the signaling process is
stochastic. The aggregate response of several thousand of these
microdomains yields a whole-cell response which dictates the cell
behavior. Here, we study analytically the statistical properties of
a population of these microdomains in response to a trigger signal.
We apply these results to understand the relationship between Ca
influx and Ca release in cardiac cells. In this context, we use a
first passage time approach to show analytically how Ca release in
the whole cell depends on the single channel kinetics of Ca channels
and the properties of microdomains. Using these results, we explain
the underlying mechanism for the graded relationship between Ca
influx and Ca release in cardiac cells.

\end{abstract}

\maketitle

The use of Ca-sensitive receptors is ubiquitous in the design of
signal transduction processes.  A basic feature of this
design is the close positioning of Ca-permeable channels to
Ca-sensitive receptors in localized regions of the cell, often
referred to as Ca microdomains~\cite{Berridge06}. Typically, the
signaling takes places when a ``trigger'' channel delivers Ca into
the microdomain, and the rise of the local Ca concentration is
sensed by Ca-sensitive receptor channels.  In this way, a trigger
input can induce a large response signal which then facilitates
downstream cellular processes. In cardiac cells this signaling
architecture is used to mediate the interaction between membrane
voltage and cell contraction~\cite{Bers06}, while in neurons it
modulates synaptic function \cite{Sharma03}.

Within a microdomain the relationship between the trigger signal and
receptor response is nonlinear, since (i) Ca receptors themselves
control the flow of a high concentration of Ca sequestered inside
the cell, and (ii) Ca receptors are clustered so that Ca released
from one receptor can induce a neighboring receptor within the
microdomain to open.   In this way a small increase in trigger
current can induce a large autocatalytic response from a cluster of
receptors, leading to an ``all or none'' excitability. However, the
spatial separation between microdomains ensures that the Ca released
in one domain does not necessarily induce release in a neighboring domain. Thus,
the cell's response to a signal such as a change in membrane
potential will be dictated by the aggregate response of a population
of independent microdomains. This feature of Ca signaling, referred
to in cardiac physiology as ``local control''~\cite{Stern92,
Wier94}, allows the cell to smoothly control its whole-cell
response, despite the all-or-none nature of Ca signaling at the
channel level.
\begin{figure}[h]
\includegraphics[width=6.5cm, height=5.6 cm]
{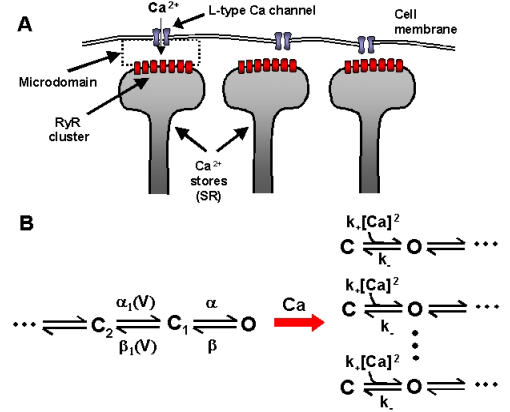} \caption{A. Illustration of Ca-mediated signaling between
LCC and RyR channels in cardiac cells. Ca is injected into the cell
via the LCCs and triggers the opening of RyR channels in the
immediate vicinity.
 B. Markov state models for LCC (left) and RyRs (right).
The vertical dots indicate many
($\sim 50-300$) RyRs; horizontal dots represent possible deeper states.}
 \label{fig1}\vspace{-3mm}
\end{figure}

Detailed simulation studies have shown that the ensemble behavior of
a population of microdomains can reproduce experimentally-measured whole-cell
currents~\cite{Greenstein06, Soeller04}. However, these studies did
not provide a concise analytic relationship between channel
statistics and whole-cell behavior.  Also, Ca receptor clusters have
been studied extensively~\cite{Shuai02, Hinch04, Bar00}, but without
directly addressing the relationship between trigger and response.
In this paper we use a first passage time approach to study the
nonlinear stochastic interaction of Ca-sensitive receptor channels
and nearby ``trigger'' channels.   We then compute the aggregate
response of a population of microdomains to show how the whole-cell
response depends on local Ca signaling.  We apply these results to
explain experimentally known features of Ca signaling in cardiac
cells.  For the sake of concreteness we have developed our analysis
within the context of Ca signaling in cardiac cells. However, the
signaling architecture we consider is found in many cellular systems
\cite{Berridge06} and should be applicable in a broader context.

The basic architecture of Ca signaling in cardiac cells is shown
schematically in Fig.~\ref{fig1}A.  Here, the signaling occurs
between voltage-sensitive L-type Ca channels (LCC) on the cell
membrane and a cluster of Ca-sensitive Ryanodine receptors (RyR)
\cite{Bers06}. The RyRs gate the flow of Ca from an internal store,
the sarcoplasmic reticulum (SR), which has a concentration roughly
$10^{4}$ times greater than in the cell.  Within the microdomain, signaling
occurs via the interaction of a few LCCs and a tightly knit cluster
of 50 to 300 RyRs. The microdomain is shaped like a pill box of
height $\sim 10$ nm and diameter $\sim 100$ nm; a typical cardiac
cell has $\sim 10^4$ microdomains distributed throughout the cell.

To model ion channels within the microdomain we use a Markov state
approach as illustrated in Fig.~\ref{fig1}B.   The single channel
properties of LCCs have been studied extensively~\cite{cavalie};
here we model a single LCC using two closed states which can
transition to a Ca-permeable open state \fnI. The membrane voltage
(V) dependence of LCCs appears in the transition rates between
closed states ($\alpha_1(V)$, $\beta_1(V)$), while transition rates
to and from the open state ($\alpha$ and $\beta$) are voltage
independent.  The activation kinetics of RyRs are also well known
\cite{zahradnikova}, and it is believed that the open probability is
regulated by several Ca binding sites acting cooperatively on the
receptor. We use a minimal model which incorporates these essential
features by making the open rate $k_{+}c^2$ (where $c$ denotes the
local Ca concentration) and closing rate $k_{-}$.  The interaction
between LCCs and RyRs is dominated by the activation kinetics, and
so it is safe to neglect deeper states in the Markov scheme that
describe slower inactivation processes.

When a cardiac cell is stimulated the rise of the membrane voltage
leads to a dramatic increase in the open probability of LCCs.  The
subsequent Ca entry into the cell induces Ca release via RyR
receptors, leading to a global rise of Ca in the cell. High
resolution optical imaging of Ca~\cite{Cleemann98} in cardiac cells
indeed reveals local elevations in Ca, referred to as Ca ``sparks".
These sparks represent the release of Ca from the SR through RyR
clusters which are activated by local LCCs.   Ca sparks have
relatively constant amplitude and duration~\cite{Cleemann98}, and
so the release flux of Ca from the SR is largely dictated by the
number of Ca sparks recruited in the cell~\cite{Shiferaw03}.  If we
denote $\Delta N_S$ to be the number of new sparks recruited during
a small time interval $\Delta t$, then
\ar \Delta N_S=N_d P_{C_1} \alpha \Delta t  P_{S} \label{sparkrate}
\br
where $N_d$ is the number of microdomains in the cell;
$P_{C_1}\alpha \Delta t$ is the probability that an LCC makes a
transition from $C_1$ to $O$ in the time interval $\Delta t$, with
$P_{C_1}$ being the probability of the closed state that directly
transitions to the open state of an LCC; and $P_{S}$ denotes the
probability that the LCC opening will trigger a spark. Thus, the key
quantity relating the local channel interactions to the whole-cell
response is the rate of spark recruitment $R_S=\Delta N_S/\Delta t$,
which depends on the trigger-response interaction described by
$P_{S}$.

To compute $P_S$ we first note that Ca ions inside the cell diffuse
at $\sim 150-300\ \mathrm{\mu m^2/s}$, and thus equilibrate over the scale of
the microdomain much faster ($\sim 0.01$ms) than the typical channel
transition times ($\sim 1$ms). Fast diffusion yields $c \approx c_o
+ \left( \tau/v \right)  \left( n \cdot i_{RyR} + k \cdot
\ica\right)$,  where $c_o$ is the resting Ca concentration outside
of the microdomain of volume $v$; $\tau$ is the time constant of
diffusion out of the microdomain; $n$ and $k$ are the number of open
RyR and LCC channels, respectively; and $i_{RyR}$ and $\ica$ are the
fluxes, in ions per unit time, through the respective single open
channels. We set $i_{RyR}=gc_{sr}$, valid for small $c$, where $g$
is the single RyR channel conductance and $c_{sr}$ is the Ca
concentration in the SR.  Since there are only a few LCC channels in
the microdomain, each with small open probability \cite{cavalie}, we
will simplify the system further and assume that there is only one
LCC channel in each microdomain so that $k=0$ or $1$.

The state of the RyR cluster can be described using a master
equation for the probability $P(n,t)$ that $n$ out of $N$ total RyR
channels in the cluster are open, given by
\begin{align}
\nonumber \frac{dP(n,t)}{dt} &= r_+(n-1)P(n-1,t) + r_-(n+1)P(n+1,t) \\
  &- \left[r_+(n) + r_-(n)\right]P(n,t)
\label{master-equation}
\end{align}
with forward transition rate $r_+(n)=\kp(N-n)c^2$ (a third-order
polynomial in $n$) and backward transition rate $r_-(n)=\km n$.  For
$N$ large we can use well-known approximation
methods~\cite{Gardiner90} to reduce this one-step process to its
corresponding Fokker-Planck equation (FPE), which we can write, in
terms of the fraction $x=n/N$ of open channels, as \ar
\frac{\partial p(x,t)}{\partial t} =-\frac{\partial}{\partial x}
\left[f(x) p(x,t)\right]  + \frac{1}{2N}\frac{\partial^2}{\partial
x^2} \left[h(x) p(x,t)\right] , \label{fp2} \br with drift
cofficient $f(x)=(r_+(Nx)-r_-(Nx))/N$ and diffusion coefficient
$h(x)/N=(r_+(Nx)+r_-(Nx))/N^2$. The diffusion term scales as $1/N$
implying the increasingly deterministic nature with larger cluster
size \cite{Fox94}.

\begin{figure}
\includegraphics[width=5.20cm, height=3.40 cm]
{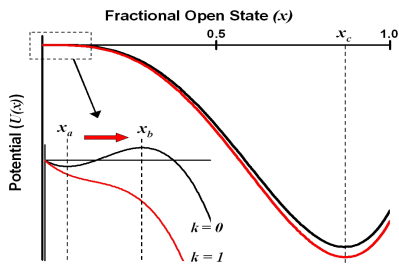} \vspace{5mm}\caption{\textit{Main:} Effective potential
landscape $U(x)$ corresponding to closed ($k=0$) or open ($k=1$)
LCC. \textit{Inset:} Barrier near origin disappears when $\ica$ is
large enough. Parameters here are $\ica=0.06\,\mathrm{pA}$,
$g=0.520\,\mathrm{\mu m^3\,\mu s^{-1}}$ and $c_o=5\,\mu M$.}
 \label{fig2}\vspace{-3mm}
\end{figure}

The equivalent Langevin description corresponding to this FPE
implies time evolution of the mean open fraction $\langle x\rangle$
dictated by the equation $\frac{d\langle x\rangle}{dt}=f(\langle
x\rangle)$. Therefore $f(x)=0$ determines the stationary points of
$x$. The deterministic component of the dynamics can be
conveniently visualized in terms of a potential $U(x)=-\int^x_0
f(x')dx'$ (Fig.~\ref{fig2}). In the case when the Ca channel is
closed ($k=0$), the potential landscape reveals a bistable system
with three fixed points. The stable point $x_a$ corresponds to an
inactive cluster with very few channels open. A second stable point
$x_c$ denotes a nearly fully open cluster. Finally, an unstable
point at position $x_b$ indicates the peak of a potential barrier
separating the closed and open cluster states. Since $x$ is small
during initial spark activation, to second order we have $f(x)
\approx \sigma +\mu x +\kp q^2 x^2$ where $\sigma=\kp s^2$ and
$\mu=-\km +2\kp q s$, with $q=N(\tau/v)gc_{sr}$ and
$s=c_o+(\tau/v)k\ica$. The equilibrium states can then be found as
$x_a \approx (\kp/\km)c_o^2$ and $x_b \approx (\km/\kp)q^{-2}$.

The interaction between the LCC and the RyR cluster is governed by
the degree to which an LCC opening will induce a transition from
$x_a$ to $x_b$.  When the LCC channel is open the barrier is reduced
(or eliminated) as shown in Fig.~\ref{fig2}, owing to the elevated Ca
concentration in the microdomain due to Ca influx via the open LCC.
If the LCC remains open until $x$ crosses $x_b$, then a spark will
almost certainly occur as $x$ proceeds to $x_c$.  On the other hand
if the channel closes before such crossing has occurred, then as the
barrier returns the cluster state will most likely return to $x_a$.
Thus, given a passage time $t$ from $x_a$ to $x_b$, the LCC open
time, which follows an exponential distribution $f_O(t')=\beta
\exp(-\beta t')$, must exceed $t$; this yields
\ar P_S(\beta,x_a,x_b)&=&\int_0^\infty dt\ e^{-\beta t}
P(x_a,x_b;t), \label{prob} \br
where $P(x_a,x_b;t)$ is the first passage time density (FPTD) from
$x_a$ to $x_b$ for the system given by Eq.~(\ref{fp2}).

The statistics of this FPTD have been worked out in detail in a
classic paper by Darling and Siegert~\cite{Darling53}, where they
show that the Laplace transform of the FPTD can be found by solving
the adjoint FPE. Linearizing the drift and diffusion coefficients so that
$f(x)=\sigma+\mu x$ and $h(x)=\sigma+\gamma x$, where $\gamma=\km+2
\kp qs$, the adjoint FPE can be written as a confluent
hypergeometric differential equation, which has the linearly
independent solutions $u_1(x)=M(m_1,m_2,y(x))$ and
$u_2(x)={y(x)}^{1-m_2}M(1+m_1-m_2,2-m_2,y(x))$, where $M$ denotes
Kummer's function~\cite{a+s} with arguments $m_1=-\beta/\mu$,
$m_2=2N\sigma(\gamma-\mu)/\gamma^2$, and $y(x)=-2N\mu(\sigma+\gamma
x)/\gamma^2$.  Imposing reflecting boundary conditions at $x=0$,
since the master equation (Eq.~(\ref{master-equation})) does not
admit negative valued states, the probability of sparking can be
written~\cite{Darling53} as \ar P_{S}(\beta,x_a,x_b)=
\frac{u_1(x_a)/u'_1(0) - u_2(x_a)/u'_2(0)}{u_1(x_b)/u'_1(0) -
u_2(x_b)/u'_2(0)} , \label{pspark}\br where $u'_1(0)$ and $u'_2(0)$
are the derivatives of the independent solutions evaluated at the
origin. Eq.~(\ref{pspark}) gives a full description of the sparking
probability given a trigger current $\ica$ with duration from the
distribution $f_O$.

The analytic formula for the spark rate (Eq.~(\ref{pspark})) can be
simplified further in the limit of large or small LCC current
($\ica$).  Using the asymptotic form of the Kummer function
\cite{a+s}, for large $\ica$ we have \ar P_S \sim \left(
\frac{\sigma+\gamma x_a}{\sigma +\gamma x_b}\right)^{\beta/\mu} .
\br  In this limit, the response is governed by the mean open time
of the LCC, and does not strongly depend on the stochastic
properties of the RyR cluster. For small $\ica$ a barrier must be
surmounted, and stochastic fluctuations due to the $N$ RyR channels
dictate the probability for sparking. In this regime, to leading
order we have \ar P_S \sim \exp{\left( \frac{2\mu}{\gamma}N(x_b-x_a)
\right)} , \label{decay} \br showing the strong dependence on RyR
cluster size $N$.

\begin{figure}
\includegraphics[width=6.0cm, height=4.80cm]{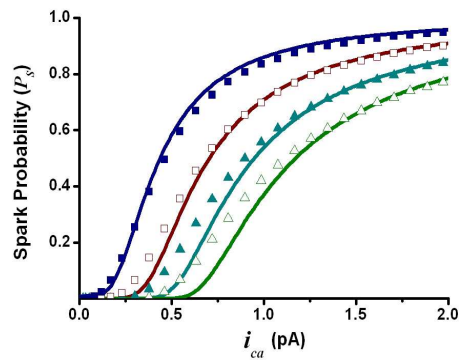}
\caption{Analytical (curves) and numerically simulated (symbols)
sparking probability $P_S$
as a function of $\ica$, conditional upon an LCC
activation, for conductances $g=0.910$ ({\tiny$\blacksquare$}),
$0.628$ ({\tiny$\square$}), $0.491\ (\blacktriangle)$, and $0.416$
({\footnotesize$\triangle$}) $\mathrm{\mu m^3\,\mu s^{-1}}$.}
\label{fig3}\vspace{-0.5cm}
\end{figure}

In order to check the validity of this analytical result we simulate
independent Monte Carlo trajectories of the master equation to
estimate the probability of sparking ($P_S$) given an LCC opening at
time $t=0$, in an ensemble of $100,000$ microdomains whose clusters
are initially closed ($x=x_a$). Within each unit a single LCC is
opened for a duration $t'$ chosen from $f_O$. We then estimate $P_S$
as the fraction of microdomains whose clusters reach $n=x_bN$ within
that time.  In Fig.~\ref{fig3} the predictions of Eq.~(\ref{pspark})
are compared with the numerical simulations.  As shown, the
agreement is quite good.
\begin{figure}
\includegraphics[width=6.0cm, height=4.3 cm]{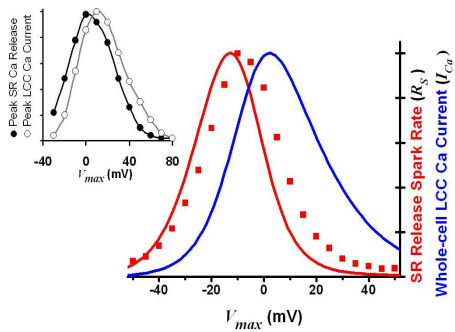}
\caption{\textit{Main:} Analytical (left curve) and numerically simulated
({\tiny$\blacksquare$}) spark recruitment rate $R_S$ as a function
of $V_{max}$ ($g=0.910\,\mathrm{\mu m^3\,\mu s^{-1}}$); whole-cell
$I_{ca}$ (right curve). \textit{Inset:} Experimental values
from~\cite{Wier94} for peak SR release current ($\bullet$) and peak
LCC current ($\circ$).  (All plots normalized to peak height)} \label{fig4}
\vspace{-3mm}
\end{figure}

In experiments, the relationship between Ca entry and Ca release can
be assessed by depolarizing the membrane to various test
voltages $V_{max}$.  The peak total currents for LCC Ca entry and SR Ca
release are then measured following the depolarization. In
Fig.~4 (inset) we reproduce the experimental data of Wier et al.
\cite{Wier94}, showing that both Ca release and Ca entry follow a
bell-shaped dependence on $V_{max}$. This relationship is well
established, and referred to as graded release.  Since the spark
recruitment rate is directly related to the amount of Ca released
from the SR, we check this relationship by computing the spark
rate $R_S$ as a function of $V_{max}$.  Figure \ref{fig4} shows the
peak spark rates due to a depolarization to $V_{max}$, using our
analytically calculated $P_S$ from Eq.~(\ref{pspark}) in
Eq.~(\ref{sparkrate}), as well as by simulating $N_d=100,000$
independent microdomains with Markov-governed LCCs initially in
state $C_2$. Numerically, the spark rates were estimated by counting
the number of sparks recruited in $1$ms intervals. Also plotted is
the peak whole-cell Ca current estimated by $I_{Ca}=N_d P_o \ica$,
where $P_o$ is the steady-state open probability of the LCC.  As
shown, consistent with graded release, the spark rate dependence on
$V_{max}$ follows a bell curve mirroring the Ca entry $I_{Ca}$, and
the analytical predictions agree semi-quantitatively with the
numerical simulations.

The graded relationship between whole-cell Ca release and Ca
entry is reflected in the spark recruitment rate
(Eq.~(\ref{sparkrate})). For large negative $V_{max}$ $\ica$ is
large and every LCC opening will trigger a spark.  In this regime
the spark rate is dominated by the voltage dependence of $P_{C1}$,
which roughly follows $P_o$, since $\alpha$ and
$\beta$ are fast voltage-independent rates. As $V_{max}$ is
increased further, $\ica$ decreases and $P_S$ is reduced as the
probability of triggering a spark is less. Similarly, $I_{Ca}$ also
decreases due to the drop in the single channel current $\ica$.
Hence, Eq.~(\ref{sparkrate}) reproduces experimentally measured
whole-cell properties in terms of the kinetics of ion channels in Ca
microdomains. A more detailed analysis of the analytic prediction
will be presented in a future publication.

In this letter we have analyzed the stochastic properties of Ca
signaling at both the whole-cell and ion channel level.  The main
result is an analytic description of the relationship between single
channel kinetics and the aggregate whole-cell response. In the
context of the cardiac cell, we have reproduced important
voltage-dependent relationships which are known experimentally, but
until now have not been explained in detail. This work should pave
the way to a more detailed understanding of Ca signaling in a wide
range of biological processes.

This work is supported by the NSF and NIH/NHLBI P01 HL078931; R.R.
thanks Tom Chou and Vladimir Minin for valuable discussions. Y.S.
thanks A. Karma for valuable discussions, and the KITP Santa
Barbara, where part of this work was completed.

\bibliography{refs}

\begin{thebibliography}{18}
\expandafter\ifx\csname natexlab\endcsname\relax\def\natexlab#1{#1}\fi
\expandafter\ifx\csname bibnamefont\endcsname\relax
  \def\bibnamefont#1{#1}\fi
\expandafter\ifx\csname bibfnamefont\endcsname\relax
  \def\bibfnamefont#1{#1}\fi
\expandafter\ifx\csname citenamefont\endcsname\relax
  \def\citenamefont#1{#1}\fi
\expandafter\ifx\csname url\endcsname\relax
  \def\url#1{\texttt{#1}}\fi
\expandafter\ifx\csname urlprefix\endcsname\relax\def\urlprefix{URL }\fi
\providecommand{\bibinfo}[2]{#2}
\providecommand{\eprint}[2][]{\url{#2}}

\bibitem[{\citenamefont{Berridge}(2006)}]{Berridge06}
\bibinfo{author}{\bibfnamefont{M.~J.} \bibnamefont{Berridge}},
  \bibinfo{journal}{Cell Calcium} \textbf{\bibinfo{volume}{40}},
  \bibinfo{pages}{405} (\bibinfo{year}{2006}).

\bibitem[{\citenamefont{Bers}(2006)}]{Bers06}
\bibinfo{author}{\bibfnamefont{D.~M.} \bibnamefont{Bers}},
  \emph{\bibinfo{title}{Excitation-Contraction Coupling and Cardiac Contractile
  Force (Developments in Cardiovascular Medicine)}}
  (\bibinfo{publisher}{Springer}, \bibinfo{year}{2006}).

\bibitem[{\citenamefont{Sharma and Vijayaraghavan}(2003)}]{Sharma03}
\bibinfo{author}{\bibfnamefont{G.}~\bibnamefont{Sharma}} \bibnamefont{and}
  \bibinfo{author}{\bibfnamefont{S.}~\bibnamefont{Vijayaraghavan}},
  \bibinfo{journal}{Neuron} \textbf{\bibinfo{volume}{38}}, \bibinfo{pages}{929}
  (\bibinfo{year}{2003}).

\bibitem[{\citenamefont{Stern}(1992)}]{Stern92}
\bibinfo{author}{\bibfnamefont{M.~D.} \bibnamefont{Stern}},
  \bibinfo{journal}{Biophys J} \textbf{\bibinfo{volume}{63}},
  \bibinfo{pages}{497} (\bibinfo{year}{1992}).

\bibitem[{\citenamefont{Wier et~al.}(1994)\citenamefont{Wier, Egan,
  Lopez-Lopez, and Balke}}]{Wier94}
\bibinfo{author}{\bibfnamefont{W.~G.} \bibnamefont{Wier}},
  \bibinfo{author}{\bibfnamefont{T.~M.} \bibnamefont{Egan}},
  \bibinfo{author}{\bibfnamefont{J.~R.} \bibnamefont{Lopez-Lopez}},
  \bibnamefont{and} \bibinfo{author}{\bibfnamefont{C.~W.} \bibnamefont{Balke}},
  \bibinfo{journal}{J Physiol} \textbf{\bibinfo{volume}{474}},
  \bibinfo{pages}{463} (\bibinfo{year}{1994}).

\bibitem[{\citenamefont{Greenstein et~al.}(2006)\citenamefont{Greenstein,
  Hinch, and Winslow}}]{Greenstein06}
\bibinfo{author}{\bibfnamefont{J.~L.} \bibnamefont{Greenstein}},
  \bibinfo{author}{\bibfnamefont{R.}~\bibnamefont{Hinch}}, \bibnamefont{and}
  \bibinfo{author}{\bibfnamefont{R.~L.} \bibnamefont{Winslow}},
  \bibinfo{journal}{Biophys J} \textbf{\bibinfo{volume}{90}},
  \bibinfo{pages}{77} (\bibinfo{year}{2006}).

\bibitem[{\citenamefont{Soeller and Cannell}(2004)}]{Soeller04}
\bibinfo{author}{\bibfnamefont{C.}~\bibnamefont{Soeller}} \bibnamefont{and}
  \bibinfo{author}{\bibfnamefont{M.~B.} \bibnamefont{Cannell}},
  \bibinfo{journal}{Prog Biophys Mol Biol} \textbf{\bibinfo{volume}{85}},
  \bibinfo{pages}{141} (\bibinfo{year}{2004}).

\bibitem[{\citenamefont{Shuai and Jung}(2002)}]{Shuai02}
\bibinfo{author}{\bibfnamefont{J.-W.} \bibnamefont{Shuai}} \bibnamefont{and}
  \bibinfo{author}{\bibfnamefont{P.}~\bibnamefont{Jung}},
  \bibinfo{journal}{Biophys J} \textbf{\bibinfo{volume}{83}},
  \bibinfo{pages}{87} (\bibinfo{year}{2002}).

\bibitem[{\citenamefont{Hinch}(2004)}]{Hinch04}
\bibinfo{author}{\bibfnamefont{R.}~\bibnamefont{Hinch}},
  \bibinfo{journal}{Biophys J} \textbf{\bibinfo{volume}{86}},
  \bibinfo{pages}{1293} (\bibinfo{year}{2004}).

\bibitem[{\citenamefont{B\'{a}r et~al.}(2000)\citenamefont{B\'{a}r, Falcke,
  Levine, and Tsimring}}]{Bar00}
\bibinfo{author}{\bibfnamefont{M.}~\bibnamefont{B\'{a}r}},
  \bibinfo{author}{\bibfnamefont{M.}~\bibnamefont{Falcke}},
  \bibinfo{author}{\bibfnamefont{H.}~\bibnamefont{Levine}}, \bibnamefont{and}
  \bibinfo{author}{\bibfnamefont{L.~S.} \bibnamefont{Tsimring}},
  \bibinfo{journal}{Phys Rev Lett} \textbf{\bibinfo{volume}{84}},
  \bibinfo{pages}{5664} (\bibinfo{year}{2000}).

\bibitem[{\citenamefont{Cavalie et~al.}(1986)\citenamefont{Cavalie, Pelzer, and
  Trautwein}}]{cavalie}
\bibinfo{author}{\bibfnamefont{A.}~\bibnamefont{Cavalie}},
  \bibinfo{author}{\bibfnamefont{D.}~\bibnamefont{Pelzer}}, \bibnamefont{and}
  \bibinfo{author}{\bibfnamefont{W.}~\bibnamefont{Trautwein}},
  \bibinfo{journal}{Pflugers Archiv} \textbf{\bibinfo{volume}{406}},
  \bibinfo{pages}{241} (\bibinfo{year}{1986}).

\bibitem[{\citenamefont{Zahradnikova et~al.}(1999)\citenamefont{Zahradnikova,
  Zahradnik, Gyorke, and Gyorke}}]{zahradnikova}
\bibinfo{author}{\bibfnamefont{A.}~\bibnamefont{Zahradnikova}},
  \bibinfo{author}{\bibfnamefont{I.}~\bibnamefont{Zahradnik}},
  \bibinfo{author}{\bibfnamefont{I.}~\bibnamefont{Gyorke}}, \bibnamefont{and}
  \bibinfo{author}{\bibfnamefont{S.}~\bibnamefont{Gyorke}},
  \bibinfo{journal}{J. Gen. Physiol.} \textbf{\bibinfo{volume}{114}},
  \bibinfo{pages}{787} (\bibinfo{year}{1999}).

\bibitem[{\citenamefont{Cleemann et~al.}(1998)\citenamefont{Cleemann, Wang, and
  Morad}}]{Cleemann98}
\bibinfo{author}{\bibfnamefont{L.}~\bibnamefont{Cleemann}},
  \bibinfo{author}{\bibfnamefont{W.}~\bibnamefont{Wang}}, \bibnamefont{and}
  \bibinfo{author}{\bibfnamefont{M.}~\bibnamefont{Morad}},
  \bibinfo{journal}{Proc Natl Acad Sci U S A} \textbf{\bibinfo{volume}{95}},
  \bibinfo{pages}{10984} (\bibinfo{year}{1998}).

\bibitem[{\citenamefont{Shiferaw et~al.}(2003)\citenamefont{Shiferaw, Watanabe,
  Garfinkel, Weiss, and Karma}}]{Shiferaw03}
\bibinfo{author}{\bibfnamefont{Y.}~\bibnamefont{Shiferaw}},
  \bibinfo{author}{\bibfnamefont{M.~A.} \bibnamefont{Watanabe}},
  \bibinfo{author}{\bibfnamefont{A.}~\bibnamefont{Garfinkel}},
  \bibinfo{author}{\bibfnamefont{J.~N.} \bibnamefont{Weiss}}, \bibnamefont{and}
  \bibinfo{author}{\bibfnamefont{A.}~\bibnamefont{Karma}},
  \bibinfo{journal}{Biophys J} \textbf{\bibinfo{volume}{85}},
  \bibinfo{pages}{3666} (\bibinfo{year}{2003}).

\bibitem[{\citenamefont{Gardiner}(1990)}]{Gardiner90}
\bibinfo{author}{\bibfnamefont{C.~W.} \bibnamefont{Gardiner}},
  \emph{\bibinfo{title}{Handbook of Stochastic Methods}}
  (\bibinfo{publisher}{Springer-Verlag}, \bibinfo{address}{Berlin},
  \bibinfo{year}{1990}).

\bibitem[{\citenamefont{Fox and Lu}(1994)}]{Fox94}
\bibinfo{author}{\bibfnamefont{R.~F.} \bibnamefont{Fox}} \bibnamefont{and}
  \bibinfo{author}{\bibfnamefont{Y.}~\bibnamefont{Lu}}, \bibinfo{journal}{Phys.
  Rev. E} \textbf{\bibinfo{volume}{49}}, \bibinfo{pages}{3421}
  (\bibinfo{year}{1994}).

\bibitem[{\citenamefont{Darling and Siegert}(1953)}]{Darling53}
\bibinfo{author}{\bibfnamefont{D.~A.} \bibnamefont{Darling}} \bibnamefont{and}
  \bibinfo{author}{\bibfnamefont{A.~J.~F.} \bibnamefont{Siegert}},
  \bibinfo{journal}{Ann. Math. Stat.} \textbf{\bibinfo{volume}{24}},
  \bibinfo{pages}{624} (\bibinfo{year}{1953}).

\bibitem[{\citenamefont{Abramowitz and Stegun}(1964)}]{a+s}
\bibinfo{author}{\bibfnamefont{M.}~\bibnamefont{Abramowitz}} \bibnamefont{and}
  \bibinfo{author}{\bibfnamefont{I.}~\bibnamefont{Stegun}},
  \emph{\bibinfo{title}{Handbook of Mathematical Functions with Formulas,
  Graphs, and Mathematical Tables}} (\bibinfo{publisher}{Dover},
  \bibinfo{address}{New York}, \bibinfo{year}{1964}).

\end{thebibliography}

\end{document}